\documentclass[twocolumn,prl,aps]{revtex4}
\usepackage{graphicx}

\newcommand{\be}{\begin{equation}}
\newcommand{\ee}{\end{equation}}
\newcommand{\bea}{\begin{eqnarray}}
\newcommand{\eea}{\end{eqnarray}}

\newcommand{\p}{\partial}

\newcommand{\la}{\langle}
\newcommand{\ra}{\rangle}
\newcommand{\lb}{\left[}
\newcommand{\rb}{\right]}
\newcommand{\lp}{\left(}
\newcommand{\rp}{\right)}
\renewcommand{\phi}{\varphi}

\def\nn{\nonumber\\}

\begin{document}
\title{Optimal non-linear passage through a quantum critical point}

\author{Roman Barankov and Anatoli Polkovnikov}
\affiliation {Department of Physics, Boston University, Boston, MA 02215}

\begin{abstract}

We analyze the problem of optimal adiabatic passage through a quantum critical point. We show that to minimize the number of defects the tuning parameter should be changed as a power-law in time. The optimal power is proportional to the logarithm of the total passage time multiplied by universal critical exponents characterizing the phase transition. We support our results by the general scaling analysis and by explicit calculations for the transverse field Ising model.

\end{abstract}
\maketitle

The adiabatic quantum computation is a very appealing concept for simulating certain class of problems~\cite{farhi, regev}. The main idea is that the system is initially prepared in a ground state of some simple Hamiltonian and then the Hamiltonian is adiabatically changed to some nontrivial regime. If one changes the Hamiltonian sufficiently slowly then the system remains in the ground state and this new ground state may encode nontrivial information. For example the adiabatic evolution can be a necessary condition for realizing topological quantum computation in cold atom systems such as $p$-wave superfluids~\cite{gurarie, dassarma}. In order to get to the nontrivial topological state one needs to cross a second order phase transition between gappless and gapped $p_x+ip_y$ phases~\cite{gurarie}. Indeed it is generally much easier to cool the system in the regime where topological excitations do not exist and then transfer the system to the nontrivial phase. Similar considerations can be applied to proposals based on the Kitaev model~\cite{kitaev}. We note that this idea can be used in a broader sense than quantum computing alone. For example, one can use the same algorithm to study ground states of strongly correlated systems. The adiabatic evolution is routinely used in cold atom systems to load them into optical lattices, change interaction strength, etc.~\cite{bloch_review}.

Of course ideally it is good to prepare the system initially in the ground state and keep the evolution slow enough so the system is never excited~\cite{farhi}. However this can be a very stringent requirement. If a  sufficiently large system is cooled in the gapless regime it is virtually impossible to avoid any excitations. In addition the requirement that the evolution is completely adiabatic might lead to extremely long computational times and make the whole idea of adiabatic quantum computation unrealistic. At the same time the quantum computation can be robust to certain appropriate level of excitations and thus one should always look for a compromise between shortening the evolution time and tolerance to the number of the excited states. Likewise if one is interested in understanding many-body states using adiabatic evolution of the Hamiltonian one should compromise between the evolution time and the heating generated in the system due to non-adiabaticity.
Usually the effects of non-adiabaticity are particularly strong in low dimensional systems~\cite{pg} especially near singularities such as critical points. By now there is an extensive literature focusing on slow dynamics near critical points~\cite{adiabatic, zurek1, jacek, fubini, caneva, dutta} and even critical lines like in the Kitaev model~\cite{krishnendu1}. Typically as a measure of non-adiabaticity one defines the density of produced defects during
the quench. Such measure is appropriate for integrable models with well defined excitations. More generally one can look into the excess energy density added to the system during the quench and come to similar conclusions~\cite{pg, note}.

Generically if one crosses a phase transition, the tuning parameter $g$ will change linearly in time near the critical point: $g\sim t/T$, where $T$ is the characteristic passage time. For convenience we assume that $g$ vanishes right at the critical point. If one is interested in minimizing the heating in the system or the number of produced defects in the system then such a linear change might not be optimal. Indeed it is intuitively clear that one should slow down precisely near the critical point and perhaps change the tuning parameter as a power law in time: $g\sim (t/T)^r$. Now one can ask the question: what is the optimal value of $r$? Is it determined by some universal properties of the phase transition? Or perhaps there is an optimal value of the power, e.g. $r\to\infty$, which is good irrespective of the details of the critical point. The answer to these questions is the main purpose of the present letter.

Let us first argue that there should be an optimal value of $r$. Indeed if $r\to 0$ then the tuning parameter
abruptly changes near the critical point, and we expect many excitations to be produced in the system. On the other hand if $r\to\infty$ then the
tuning parameter changes very smoothly near the critical point but then changes very abruptly inducing a very strong heating in the system. If we want to find an optimal path at a given constraint that the total passage time $T$ is fixed we clearly have to find a compromise between being as slow as possible right at the critical point and being not very fast when the tuning parameter is finite.

Let us now be more specific and assume that
\be\label{coupling}
g(t)\approx g_0 |t/T|^r{\rm sign}(t),
\label{lam_t}
\ee
where $g_0$ is the characteristic value of the tuning parameter in the final state. Note that the time $t$ can change either within $[-T,T]$ corresponding to the crossing of the critical point or within $[0,T]$ corresponding to starting right at the critical point and changing the tuning parameter from zero to the final value $g_0$. The latter regime is realized, for example, if one loads interacting one-dimensional Bose gas into an optical lattice~\cite{claudia}. A generic quantum phase transition at $g=0$ is characterized by critical exponents~\cite{subir}. The exponents, which are relevant to us are $\nu$ and $z$. The correlation length exponent $\nu$ characterizes the divergence of the correlation length $\xi$ at small $g$: $\xi\sim g^{-\nu}$, while the dynamical critical exponent $z$ defines the relation between a characteristic spectral energy $\Delta$ and the correlation length: $\Delta\sim \xi^{-z}$, (see Ref.~\cite{subir} for more details).
Instead of $T$ it is convenient to introduce a dimensionless parameter of  adiabaticity, $\delta={1/(T\Delta_0)}$, where $\Delta_0$ corresponds to $g=g_0$. For gapped systems $\Delta_0$ is related to the lowest excitation energy in the system, while for gapless systems $\Delta_0$ is the energy scale at which the spectrum qualitatively changes from the low energy to the high energy behavior~\cite{subir}. Our main conclusion is that in the adiabatic limit $\delta\ll 1$ the optimal power $r_{\rm opt}$ depends only on $\delta$ and the universal exponents characterizing the phase transition:
\be
r_{\rm opt}\approx -{1\over z\nu}\ln\left[{\delta\over C}\ln (C/\delta)\right],
\label{main}
\ee
where $C$ is a non-universal constant of the order of unity, which depends on the details of the phase transition and precise choice of $\Delta_0$ in the definition of the parameter $\delta$.

 The result (\ref{main}) is quite remarkable. It implies that the optimal power (modestly logarithmically) depends on the total passage time. It approaches infinity only for infinitesimally slow processes. If the passage time is large but finite then $r_{\rm opt}$ is usually a relatively small power of the order of unity. Interestingly dimensionality does not enter the result (\ref{main}).
Similarly, we get the result for the density of the produced defects if one uses the optimal adiabatic passage:
\be
n_{\rm opt}\sim \left[{\delta\over C}\ln(C/\delta)\right]^{d/z}.
\label{main1}
\ee
Here $d$ is the dimensionality of the system. In dimensions higher than one these results immediately generalize to the situation when instead of critical point one has a critical surface of dimension $d_{\parallel}$. E.g. $d_\parallel=1$ for the Kitaev model. Then in Eq.~(\ref{main1}) one has to use $d_\perp=d-d_\parallel$ instead of $d$~\cite{krishnendu1}. The scaling of the density of the produced defects for optimal passage is quite different from what one would get for a generic linear in time crossing of the critical point~\cite{adiabatic}: $n_{\rm lin}\sim \delta^{d\nu/(z\nu+1)}$. It is easy to check that:
\be
{n_{\rm opt}\over n_{\rm lin}}\sim \left[\left({\delta\over C}\right)^{1\over z\nu+1}\ln(C/\delta)\right]^{d/z}\ll 1.
\ee
so that optimal passage gives a significant improvement over the linear one. We note that the constant $C$ can always be reabsorbed in the precise definition of $\delta$. To simplify the notations we will assume this is the case and we skip $C$ in the expressions below.

We note that the optimal path does not have to be a power law function at $t\to 0$ as it is assumed in Eq.(\ref{coupling}). However, the power law behavior is very universal and is expected to be valid close to the critical point. The specific details of the optimal functional dependence on $t$ away from the critical point are likely to be non-universal. However, since the defects are mainly produced at $g(t)$ close to zero, we do not expect that our results will be qualitatively affected by the deviations from the power law. In the rest of the paper we will sketch the derivation of our main result and consider a specific example of the transverse field Ising model.

To calculate the number of the defects produced during the process we will use the Fermi Golden rule analysis in the adiabaticity parameter $\delta$. As it was shown in  Refs.~\cite{pg, adiabatic} in the case of linear quench unless the non-adiabtic regime is realized~\cite{pg} this analysis gives the correct dependence of the density of the defects on $\delta$ but does not reproduce the prefactor correctly. For example, the Fermi Golden rule calculation overestimates the prefactor by approximately two times for the transverse Ising model. Technically the mistake comes from the fact that at very small energies the process is nonadiabatic and the Fermi Golden rule is not precise. However, one can show, that unless the system is in the non-adiabatic regime the mistake reduces to a mere factor, which is always bounded between one and another constant of the order of one. This result is true for any choice of time dependence of the tuning parameter $g$ thus the general scaling results based on the Fermi Golden rule analysis are correct. We will illustrate the difference between Fermi Golden rule and exact approaches below using a particular example of the transverse field Ising model. As we will show there is only a minor difference between them.

 Within the Fermi Golden rule the density of the excited states is given by the following expression:~\cite{adiabatic}
\be\label{eq:exc}
n=\sum_{p\ne 0} \left| \int_{-\infty}^{\infty} dt \la p,t| \p_t| 0,t \ra  e^{i\int_0^{t} dt'\lb \omega_p(t')-\omega_0(t')\rb} \right|^2,
\ee
where the last summation is over the $d$-dimensional phase space. Here, $|p,t\ra$ is the instantaneous eigen-state of the Hamiltonian describing the system, and $\omega_p(t)$ is the excitation energy. We can next use general scaling arguments (see also Ref.~\cite{adiabatic}) to find the dependence of $n$ on the rate $\delta$ and the exponent $r$. Thus for the excitation spectrum we can use
\be
\omega_p-\omega_0=p^zf(\Delta/p^z)=p^zf\lp|\delta t|^{r z\nu}/p^z\rp,
\ee
where $f(x)$ is a scaling function satisfying $f(x)\sim 1$ at $x\ll 1$. For gapped systems we should also have $f(x)\sim x$ at $x\gg 1$. The latter condition implies that $\omega_p-\omega_0\sim \Delta$ for $p\to 0$. If the system is gapless then the asymptotics of $f$ will be somewhat different and our analysis should be modified accordingly.

Dependence of the gap on time, $\Delta=|t\delta|^{rz\nu}$ suggests rescaling the variables
\be\label{eq:rescaled_variables}
\tau=\frac{t\,\delta^{\frac{rz\nu}{rz\nu+1}}}{(rz\nu+1)^{\frac{1}{rz\nu+1}}},\quad q=p\lb\delta (rz\nu+1)\rb^{-\frac{r\nu}{rz\nu+1}},
\ee
which leads to the following expression for the phase defined as $\varphi_q=\int_0^t(\omega_p(t')-\omega_0(t'))dt'$:
\be\label{eq:phase_gen}
\varphi_q(x)= \lp 1+\frac{1}{rz\nu}\rp q^{z(1+\frac{1}{rz\nu})}\int_0^x dy\,y^{\frac{1}{rz\nu}-1}\,f(y),
\ee
where $x(t)={\rm sign}(\tau)|\tau|^{rz\nu}/q^z$. Similarly one can rescale the matrix element in Eq.(\ref{eq:exc}),
$dt\la p,t|\p_t|0,t\ra = dxG(x)$,
in which $G(x)\approx{\rm const}$ at $|x|\ll 1$.

The main contribution to the number of defects in Eq.(\ref{eq:exc}) can be extracted from the behavior of the phase factor in Eq.(\ref{eq:phase_gen}), which is controlled by the asymptotics of the scaling function $f(x)$. At $|x|\ll 1$, the phase can be approximated by $\varphi_q(x)\approx q^{z(1+\frac{1}{rz\nu})}(rz\nu+1) x^{1/rz\nu}$, while at $|x|\gg 1$ it is $\varphi_q(x)\approx (q^zx)^{(1+\frac{1}{rz\nu})}$ approaching a linear function at $r\gtrsim 1$. The latter asymptotics nearly independent of the loading power corresponds to the dominant contribution of the excited states to the number of defects.
The characteristic rescaled momentum $q_*\simeq 1$ at which the transitions become suppressed provides an estimate for the width of excited energy states $p_*=(r\delta)^{r\nu/(r z\nu+1)}$.

Combining the scalings above we find
\bea
n\sim \lp r\delta\rp^{dr\nu/(r z\nu+1)},
\label{n}
\eea
We note that the scaling of $n$ with $\delta$ for an arbitrary power $r$ was recently obtained in Ref.~\cite{krishnendu2} using similar methods. He we generalized that result to include the dependence of the scaling on power $r$, which is crucial for finding the optimum. Indeed it is this dependence, which encodes non-adiabatic transitions away from the critical point. Let us point that Eq.~(\ref{n}) agrees with our general expectations outlined in the beginning of the text. Thus at small $r$ the density of excitations is large because the exponent of $\delta$ is small. On the other hand for $rz\nu\gg 1$ the exponent of $\delta$ saturates at a constant value and does not depend on $r$ any more while the prefactor grows with $r$. Clearly for any given $\delta$ there should be an optimal value of $r$. And indeed if one extremizes Eq.~(\ref{n}) with respect to $r$ one finds that there is a minimum of $n$ occurring at
\be
r_{\rm opt}\delta\approx \mathrm e^{-r_{\rm opt}z\nu}.
\ee
At $\delta\ll 1$ this equation results in Eqs.~(\ref{main}) and (\ref{main1}).

\begin{figure}[t]
\includegraphics[width=3.5in]{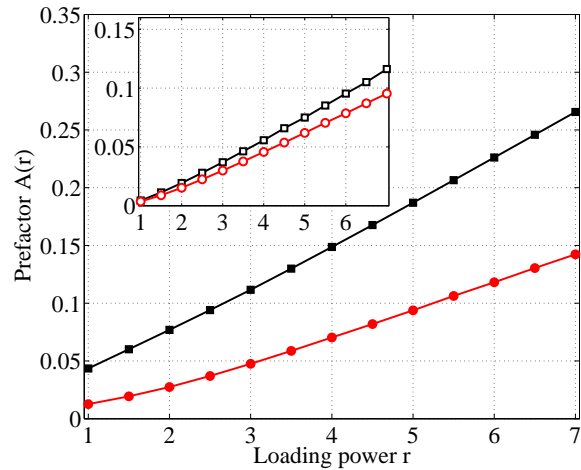}
%
\caption[]{The prefactor $A(r)$ in the number of defects~Eq.~(\ref{eq:exc_prob_TG}) as a function of the loading power for the passage through the critical point in the transversed field Ising model: the exact calculation (red solid line) and perturbative analysis (black solid line) of the adiabatic passage. {\it Inset:} prefactor $A(r)$ for the loading.}
\label{fig:pre_ising}
\end{figure}

To illustrate our results we will use a specific one-dimensional transverse field Ising model defined by the Hamiltonian:
\be
\mathcal H=\sum_{i} \sigma^z_i\sigma^z_{i+1}+(1-g)\sum_i \sigma^x_i,
\label{tfim}
\ee
where $\sigma^z_i$ and $\sigma^x_i$ denote Pauli matrices. This model describes a second order phase transition between two gapped phases at $g=0$~\cite{subir} (we assume that $g\leq 1$). The critical indices corresponding to this phase transition are $\nu=z=1$. We will consider two generic possibilities: ``a passage'', where one starts deeply in the one phase and changes $g$ through the critical point and ``loading'' where one starts exactly at the quantum critical point $g=0$ and increases the tuning parameter to drive the system into one of the phases. The second possibility, for example, naturally arises if one loads commensurate bosons to a one-dimensional periodic potential~\cite{claudia}. The problem maps exactly to the transverse field Ising model with the amplitude of the lattice potential being proportional to $1-g$.

Using the Jordan-Wigner transformation one can map the Hamiltonian (\ref{tfim}) to the Hamiltonian of noninteracting fermions~\cite{subir}:
\be
\mathcal H=\sum_i \left(c_i^\dagger c_{i+1}+c_{i+1}^\dagger c_i^\dagger+(g-1)c_i^\dagger c_i+h.c.\right).
\ee
This Hamiltonian can be diagonalized by standard Bogoliubov transformation. Then the excitations correspond to the pairs of fermions moving with opposite momenta. Near the critical point the excitation energy of a pair and the matrix elements are given by~\cite{adiabatic}
\be
\omega_p\approx 4\sqrt{p^2+g^2},
\label{en}
\ee
and
\be
\la p,g| \p_g| 0,g \ra\approx {i\over 2}{p\over p^2+g^2}.
\label{m_el}
\ee
The factor of four in the equation for the spectrum comes from the fact that the quasiparticles are excited in pairs with opposite momenta.
The matrix element (\ref{m_el}) corresponds to the pair of excitations with momenta $p$ and $-p$. We assume that $g=g_0(t/T)^r{\rm sign}(t)$ and $t\in[-T,T]$. We note that for large $T$ the number of excitations does not depend on the upper limit $g_0$ so we will choose $g_0=1$ and then $\delta=1/(Tg_0)=1/T$. Using Eqs.~(\ref{en}), (\ref{m_el}), and (\ref{eq:exc}) it is straightforward to show that within the Fermi Golden Rule the expression for the density of excitations at small values of $\delta$ assumes the form:
\be\label{eq:exc_prob_TG}
n\approx \lb A(r)\delta\rb^{r/(r+1)},
\ee
where within the perturbation theory
\bea\label{eq:phase}
A(r)&=&(r+1)\lb\int_{0}^\infty\frac{dq}{2\pi}\left|\int_{0}^\infty \frac{dx}{1+x^2}\cos\varphi_{q}(x)\right|^2\rb^{^{1+\frac{1}{r}}}\nn
\varphi_{q}(x)&=&4\lp 1+\frac{1}{r}\rp q^{1+\frac{1}{r}}\int_0^{x} dy\,y^{\frac{1}{r}-1}\,\sqrt{1+y^2}.
\eea
The phase $\varphi_{qr}(x)$ at $r\gtrsim 1$ in Eq.(\ref{eq:phase}) is a function of a rescaled momentum $q$ which reaches the order of unity at a characteristic momentum $q_*=r^{r/(r+1)}$. By evaluating the integral in Eq.~(\ref{eq:exc}) one obtains the scaling (\ref{eq:exc_prob_TG}) with
with the linear function $A_{per}(r)\approx 0.039 (r - 0.2)$ for the passage  and $A_{per}(r)\approx 0.02 (r-0.12)$ for the loading~\cite{comment}.  For the transverse field Ising model the density of the produced defects can be also found exactly since for each momentum one needs to solve an independent Landau-Zener problem~\cite{jacek}. The result of this calculation reproduces the dependence (\ref{eq:exc_prob_TG}) with the following asymptotical behavior of $A_{ex}(r)$ at large $r$: $A_{ex}(r)\approx 0.025(r-1.3)$ for the passage and $A_{ex}(r)\approx 0.017(r-1.3)$ for the loading. The linear scaling of $A(r)$ with $r$ is in complete agreement with general argument presented above if one sets $z=\nu=1$.

The exact prefactors $A(r)$ are compared in Fig.\ref{fig:pre_ising} with approximate ones obtained from the perturbative calculation in Eq.(\ref{eq:phase}). In all cases $A(r)$ linearly increases with $r$ at large $r$. As we mentioned earlier the ratio of the exact and golden rule result is a number which only slightly depends on $r$ and saturates at a constant independent of $r$ at large $r$.

\begin{figure}[t]
\includegraphics[width=3.5in]{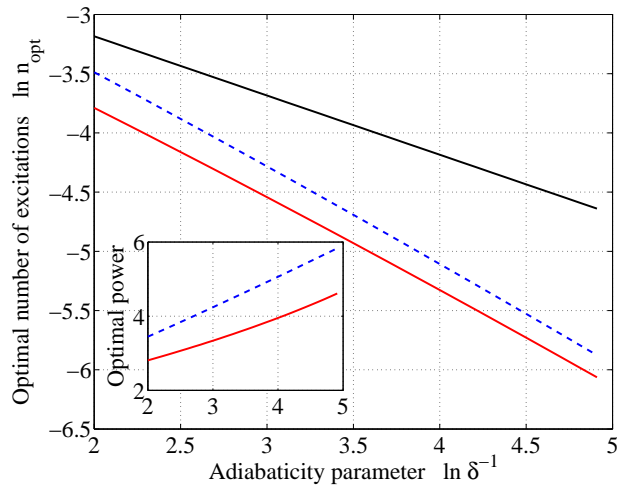}
%
\caption[]{The optimal number of excitation as a function of the adiabaticity parameter for the passage through the critical point in the transversed field Ising model. The result of the exact numerical calculation (red solid line) is compared with analytic estimate (dashed blue line) and linear passage (black solid line). {\it Inset:} the optimal loading power as a function of adiabaticity parameter for the passage through the critical point in the transversed field Ising model: the exact calculation (solid red line) and analytic estimate (dashed blue line) in Eq.(\ref{main_ising}).}
\label{fig:opt_number_comb}
\end{figure}

Minimizing the density of excitations with respect to the power $r$, one arrives at the optimal power of loading:
\be
r_{\rm opt}\approx -\ln\lb \frac{\delta}{C}\ln\lp C/\delta\rp\rb,
\label{main_ising}
\ee
where for the passage constant $C=14.7$ is extracted from the linear fit of the prefactor $A_{ex}(r)$, and similarly $C=21.6$ for the loading.
It is a weak function of the dimensionless loading parameter $\delta$. This equation agrees with our general result (\ref{main}). The analytic expression for the optimal loading power gives the right asymptotic behavior at small values of the adiabaticity parameter as shown in Fig.\ref{fig:opt_number_comb}. The dependence of the optimal number of excitations on the adiabaticity parameter is shown in Fig.\ref{fig:opt_number_comb}. At small values of $\delta$, Eq.(\ref{main1}) is a good approximation of the exact result. The optimal passage gives significantly less defects than the linear one.

In conclusion we showed that there is a universal optimum adiabatic path which minimizes the number of excitation produced in the system if the latter is slowly driven through a quantum phase transitions. This path is characterized by a power law dependence of the tuning parameter near the critical point, where the power is proportional to the logarithm of the total passage time multiplied by universal critical exponents characterizing the phase transition. Our results can be used both for optimization of adiabatic quantum computation algorithms and for minimizing heating in cold atom systems in strongly correlated regimes.

{\em Acknowledgements} This work was supported by AFOSR YIP.


\begin{thebibliography}{99}
\bibitem{note} Note that in gapped systems density of defects and the energy density are proportional to each other. However, if the final state is gapless than the scaling of the two quantities might be different (see Ref.~\cite{pg}).
\bibitem{comment} For the loading, in Eq.~(\ref{eq:phase}) the phase factor $\cos\varphi_q$ is substituted by $(1/2)\exp(i\varphi_q)$.
\end{thebibliography}
\end{document}